\documentclass[12pt,a4paper]{article}
\usepackage{t1enc}
\usepackage[latin1]{inputenc}
\usepackage[english]{babel}
\usepackage{graphics}
\begin{document}

\title{NONLINEAR EFFECTS IN CW MODE-LOCKED SOLID-STATE LASERS WITH
SEMICONDUCTOR SATURABLE ABSORBERS}
\author{V. L. Kalashnikov, D. O. Krimer}
\maketitle

\begin{abstract}
The influence of nonlinear properties of semiconductor saturable absorbers on
ultrashort pulse generation was investigated. It was shown, that linewidth enhancement,
quadratic and linear ac Stark effect contribute essentially to the mode locking in
cw solid-state lasers, that can increase the pulse stability, decrease pulse duration and 
reduce the mode locking threshold.
\end{abstract}

\begin{center}
International Laser Center, 65 Skorina Ave., Bldg. 17, Minsk 220027, BELARUS%
\\[0pt]
Tel/Fax: /375-0172/ 326-286, E-mail: vkal@ilc.unibel.by,

http://www.geocities.com/optomaplev
\end{center}

\section{Introduction}

Recently, a considerable progress has been made in self-starting femtosecond
lasers using semiconductor structures [1]. This allows to generate the
pulses with duration of 5.5 fs directly from the resonator [2]. Laser
systems with semiconductor modulators combine the advantages of the
Kerr-lens mode locking system with self-starting operation and alignment
insensitivity [3]. The most striking feature of semiconductor absorbers used
in the experiments is a long ($\sim $ 100 fs $\div $ 10 ps) recovery time as
compare with pulse duration. To explain such extremely short pulse
generation the soliton mode-locking mechanism was proposed [4]. This
mechanism involves the stabilization of the Schrödinger soliton against cw
laser continuum (noise) due to noise decay within the positive net-gain
window, which is much longer than the pulse. However, not only semiconductor
loss saturation, but also others nonlinear effects, in particular,
absorption linewidth enhancement and ac Stark-effect neglected in the above
theory can contribute to mode-locking and produce a strong self-amplitude
modulation (see, for example [5]) and thus should be taken into account.

Here we present simple models for mode-locking due to the linear and
quadratic Stark shift of the excitonic resonance and ultra-short pulse
stabilization due to linewidth enhancement.

\section{Ultrashort pulse generation in the presence of the absorption
linewidth enhancement}

It is known, that the semiconductor structures used as passive modulators
possess an extremely high nonlinearity, which depends on the carrier density
and, consequently, on the pulse energy (see, for example, [6]). This
produces a strong energy-dependent self-phase modulation (SPM), which is
proportional to loss coefficient. The corresponding coefficient of
proportionality (Henry's factor) is about -3 $\div $ -8 [7, 8].

The aim of this section is to describe an ultrashort pulse generation in cw
solid-state laser in the presence of fast nonlinear refraction in active
medium and carriers density dependent SPM in semiconductor absorber. As we
shall demonstrate later on, the last factor transforms the pulse
characteristics essentially and can stabilize an ultrashort pulse against
automodulational instability. In addition, it will be shown that the
absorption linewidth enhancement produces a negative feedback, which leads
to multistable operation. One should note the difference from an existing
theory [4], where quasi-Schrödinger laser soliton was studied.

\subsection{Model}

Based on the self-consistent field theory [9] and taking into account the
gain, the saturable loss in semiconductor, frequency filtering, the GVD and
the SPM we arrive to the master equation:

\vspace{1pt}

\begin{eqnarray}
\frac{\partial a(k,t)}{\partial k} &=&{\alpha -\Gamma \lbrack \exp [-\frac{%
\epsilon }{U_{a}}]+i\chi (\exp [-\frac{\epsilon }{U_{a}}]-1)](1+\frac{%
\partial }{\partial t})^{-1}+} \\
&&[\frac{1}{(1+\frac{\partial }{\partial t})}-1]-l+ id\frac{\partial ^{2}}{%
\partial t^{2}}-i\beta \left| a(k,t)\right| ^{2}+i\phi a(k,t),  \nonumber
\end{eqnarray}

where $a(k,t)$ is the field, $k$ is the transit number, $t$ is the local
time, $\alpha $ is the dynamically saturated gain, $l$ is the linear loss, $%
\Gamma $ is the initial saturable loss, $\varepsilon $ is the instant pulse
energy flux, $U_{a}$ is the loss saturation energy flux, $\phi $ is the
phase delay after the full round trip, $d$ is the GVD coefficient, $\beta $
is the SPM coefficient of active medium, $\chi $ is the Henry's factor. The
term in the straight parentheses with the derivation stands for the
frequency filtering. All times are normalized to the inverse filter
bandwidth $t_{f}$ and the equal bandwidths for the loss and frequency filter
are assumed for the sake of simplicity. The gain dispersion is neglected.

An expansion of eq. (1) into the series in energy and account for the
dynamical loss and gain saturation by the pulse energy

$\epsilon =\int\limits_{t_{0}}^{t}\left| a(k,t\prime )\right| ^{2}dt\prime $
($t_{0}$ is the time moment corresponding to the pulse peak) yields:

\vspace{1pt}

\begin{eqnarray}
\frac{\partial a(k,t)}{\partial k} &=&{\alpha _{0}(1-\tau \epsilon +\frac{%
(\tau \epsilon )^{2}}{2})-l-\gamma _{0}-} \\
&&{\gamma _{0}(1+i\chi )(\frac{\epsilon ^{2}}{2}-\epsilon )-(1-\gamma
_{0}(1-\epsilon ))\frac{\partial }{\partial t}+}\newline
\nonumber \\
&&(1-\gamma _{0})\frac{\partial ^{2}}{\partial t^{2}}+id\frac{\partial ^{2}}{%
\partial t^{2}}-i\chi \gamma _{0}\epsilon \frac{\partial }{\partial t}+ 
\nonumber \\
&&i\phi -ip\left| a(k,t)\right| ^{2}a(k,t),  \nonumber
\end{eqnarray}

where $\alpha _{0}$ and $\gamma _{0}$ are the saturated gain and the
saturated loss at the pulse peak, respectively. In eq. (2) the pulse energy
flux is normalized to the saturation energy flux of the absorber $U_{a}$, $%
\tau $ is the ratio of the loss saturation energy flux to the gain
saturation energy flux (with account for the mode cross-sections at the
absorber and at the active medium), $p=\frac{\beta U_{a}}{t_{f}}$.

A soliton-like solution of eq. (2) is sought in the form

\vspace{1pt}

\vspace{1pt}%
\begin{equation}
a(k,t)=a_{0}sech^{1+i\psi }[(t-k\delta )/t_{p}]\exp [i\omega (t-k\delta )],
\end{equation}

where $a_{0}$ is the pulse amplitude, $\delta $ is the round-trip delay so
that $t_{0}=k\delta $, $t_{p}$ is the pulse duration, $\omega $ is the
frequency shift from the center of the transmission band of the filter, $%
\psi $\ is the pulse chirp. Substitution of the expression (3) into (2)
gives the set of six algebraic equations. Solution of this set gives
explicit expressions for six unknown $a_{0}$, $t_{p}$, $\psi $, $\omega $, $%
\delta $\ and $\phi $, however too clumsy to write them out here.

To relate the parameters of our model to those controllable in experiment we
have to recalculate the saturated loss at pulse the peak $\gamma _{0}$ with
respect to the initial saturable loss $\Gamma $ : $\gamma _{0}=e^{-E/2},$
assuming that the loss recovery time is much shorter than the cavity period $%
T_{cav}$ and much longer than the pulse duration. Here $E=\int\limits_{-%
\infty }^{\infty }\left| a\right| ^{2}dt$ is the full pulse energy. Although
in the typical femtosecond lasers the gain saturation energy is much bigger
than the loss saturation energy (in our calculations $\tau =0.0015$, which
corresponds to the absorption saturation flux of $100\;\mu J/cm^{2}$,
Ti:sapphire active medium and the beam radii of $30$ and $106\;\mu m$ at the
active crystal and at the absorber, respectively), our calculations have
shown, that the balance between these two factors noticeably affects
ultrashort pulse parameters. The saturated gain $\alpha _{0}$ can be found
as follow: $\alpha _{0}=\alpha _{m}\frac{1-\exp (-U)}{1-\exp (-U-\tau E)}%
\exp (-\tau E/2)$ , where $U$ is the pump power normalized to $\frac{\sigma
_{14}T_{cav}}{h\nu }$, $h\nu $ is the pump photon energy, $\sigma _{14}$ is
the absorption cross section of the active medium, $\alpha _{m}$ is the gain
at full inversion. This equation gives an additional relation for
determining unknown system's parameter $\alpha _{0}$.

\subsection{Ultra-short pulse characteristics and pulse stability}

Our analysis showed, that in the presence of SPM in the active medium ($%
p\neq 0$ in eq. (2)), there exist two distinctly different pulse-like
solutions of eq. (2) (parameters of this solution are presented by curves 1
of Figs. 1 - 3, where the solid lines denote a stable solution). The
difference in the nature of this two solutions is explained by contribution
from the different pulse-forming mechanisms: 1) the Schrödinger soliton
mechanism, producing chirp-free pulse with zero frequency shift and 2)
``laser'' (dissipative) mechanism, producing essentially chirped
quasi-soliton with non-zero frequency shift (curves 1 of Fig. 2). Both
solutions have a minimum in duration for a certain pump power and GVD
(curves 1 of Fig. 3) and nearly linear dependence of the pulse energy on the
pump power.

Even a small energy-dependent nonlinear refraction ($\chi \neq 0$ in eq.
(2)) «mixes» these two states, so that the chirp compensation is possible
only for relatively large pump (Fig. 1, curves 2, 3) and for the shortest
pulse the chirp remains uncompensated (Fig. 3, curves 2, 3). The features in
the behavior of the pulse parameters in the presence of the linewidth
enhancement are the broadening of the pump power region where an ultrashort
pulse generation is possible (the growth of the maximal allowable pump
power) and the increase of the Stokes shift of the pulse carrier frequency.
The last factor produces a negative feedback due to the shift of the pulse
spectrum from the gain band that decreases the pulse energy for large pump
powers and broadens the region of the pulse existence.

Now we have to investigate the stability of the solution obtained. The
stability against small perturbations of the pulse parameters (i.e.
amplitude, duration, frequency shift, chirp and energy) is considered.
Substituting a perturbed solution (3) into eq. (2) and expanding it into the
time series we get an equations set for the evolution of the pulse
parameters:

\vspace{1pt}

\vspace{1pt}%
\begin{equation}
\frac{da_{0}}{dk}=a_{0}[\alpha 0-\gamma _{0}-l-\omega ^{2}(1-\gamma
_{0})-\upsilon (1-\gamma _{0}-d\psi )],
\end{equation}

\begin{equation}
\frac{d\omega }{dk}=a_{0}^{2}[\gamma _{0}(\chi -\omega -\psi -\chi \omega
\psi )+\alpha _{0}a_{0}^{2}\psi \tau +2\upsilon \omega (\gamma \psi
^{2}+\gamma _{0}-\psi ^{2}-1)],
\end{equation}

\begin{equation}
\frac{d\upsilon }{dk}=a_{0}^{4}(\gamma _{0}-\alpha _{0}\tau
^{2})+2a_{0}^{2}\gamma _{0}\upsilon (\chi \psi -1)+2\upsilon ^{2}(3d\psi
+\psi ^{2}+\gamma _{0}-\gamma _{0}\psi -1),
\end{equation}

\begin{eqnarray}
\frac{d\psi }{dk} &=&-2a_{0}^{2}(\gamma _{0}\chi +p+\gamma _{0}\chi \psi
^{2})+\frac{a_{0}^{4}}{\upsilon }(\gamma _{0}\chi -\gamma _{0}\psi +\alpha
_{0}\tau ^{2}\chi )+ \\
&&2\upsilon (\gamma _{0}\psi ^{2}+\psi \gamma _{0}-2d\psi ^{2}-\psi
^{2}-2d),\,  \nonumber
\end{eqnarray}

where $\upsilon =\frac{1}{t_{p}^{2}}$.

To obtain the condition for the pulse stability against energy perturbation
we followed the scheme presented in [10]: integration of eq. (2) and summing
up with its complex conjugate gives the energy conservation law. From this
integral of motion the condition for the decay of the perturbation of the
pulse energy follows:

\vspace{1pt}\vspace{1pt}%
\begin{eqnarray}
\lefteqn{-\alpha _{0}\tau e^{\frac{\tau E}{2}}(1+e^{-\tau E})+l+(1+e^{-E})-}
\\
&& \frac{2\sqrt{\upsilon }}{3}(1-\gamma _{0})(1+\psi ^{2}+3\omega
^{2}/\upsilon )-\frac{2}{3}\gamma _{0}a_{0}^{2}(\chi \psi -1) <0  \nonumber
\end{eqnarray}

Here we assumed that the loss and the gain saturation obey the exponential
law, which is the case for quasi-two level system with a relaxation time
much longer than the pulse duration. Our stability analysis is a
modification of the analysis of stability against laser noise presented in
[10] with energy perturbation playing the role of gain saturation by the
noise continuum.

Negative real part of the Jacobean of the eqs. set (4 - 8) is the condition
for the pulse stability. It is seen from eqs. (4) and (8), that the
frequency shift, chirp and spectral filtering (forth term in eq. (8))
stabilize the pulse against energy and amplitude perturbations, the «slow»%
(first term in eq. (5) and second term in eq. (7)) and the «fast» (first
term in eq. (7)) SPM stabilize the pulse against frequency and chirp
perturbations. This stabilization is provided by negative feedback due to
pulse chirping and frequency shift in the presence of finite absorption and
filter bandwidths.

Condition (8) is the condition of the pulse stability against laser noise
generated within the frame of the pulse and it should be completed by the
stability analysis against the noise generated behind the pulse within the
window of the positive net-gain. As was shown in [11], the main stabilizing
factor in this case is the difference in the group velocities of the pulse
and the noise. The delay of the noise with respect to the positive net-gain
window produces an additional loss for it and, consequently, stabilizes the
pulse. A corresponding mathematical formulation for this condition leads to
evolutional equation for cw- noise with intensity $N$ assuming that the
noise consists of spontaneous spikes with durations much longer than $t_{f}$:

\vspace{1pt}\vspace{1pt}%
\begin{equation}
\frac{dN(k,t)}{dk}=\left\{ \alpha _{0}e^{-\tau E/2}-l-V-\delta \frac{%
\partial V}{\partial t}\right\} N(k,t),
\end{equation}

where $V=(1+(e^{-E}-1)e^{-t/T_{a}})$ is `` potential'' created by the pulse,
the derivative describes the time shift of the noise with respect to the
pulse. From this the condition for the decay of the noise energy results:

\begin{equation}
\alpha _{0}e^{-\frac{\tau E}{2}}-l-e^{-E}+\frac{\delta }{T_{a}}(1-e^{-E})<0.
\end{equation}

The solutions of eq. (2) which is stable against parameter perturbations
(automodulational stability) are shown in Figs. 1 - 3 by solid lines. It
should be noted, that the soliton mode locking mechanism does not work over
the full region of pulse existence, which is evidenced by essentially
non-zero chirp of the pulse and one may conclude that the pulse stability is
provided here by interplay between others lasing factors . As is seen from
Figs. 1 - 3, a relatively large nonlinear refraction in semiconductor causes
a bistable operation: there appear two stable solutions (curves 4), that
exist simultaneously for some pump power range around 4 W. One of these
solutions has a smaller chirp, larger Stokes shift, shorter duration and
lower energy.

Negative feedback due to absorption linewidth enhancement stabilizes the
pulse against pulse energy perturbation. The mechanism of such stabilization
is explained by dependence of the pulse frequency shift on the pulse energy:
the increase of the pulse energy produces a bigger frequency shift (curves
4, 5 of Fig. 2 b), which makes the pulse amplification inefficient due to
the bad overlap of the pulse and gain spectra. This prevents from further
growth of the pulse energy; the opposite situation prevents from the
decrease of the pulse energy.

Fig. 4 demonstrates the regions of the pulse existence $A$, automodulational
stability $B$, stability against laser noise $C$ and full pulse stability $D$%
. It is seen, that in the absence of linewidth enhancement there is no
automodulational stability due to insufficiency of the negative feedback in
the system (Fig. 4, a). The increase of the negative linewidth enhancement
factor increases the threshold of the pulse generation due to nonlinear loss
produced by pulse frequency shift, but the automodulational stability region
arises due to the action of «slow» SPM in the semiconductor (Fig. 4, b).
This is accompanied by the pulse shortening down to shortest possible
duration $t_{f}$ and the full pulse stabilization for some pump and GVD
(dark region $D$ of Fig. 4, b).

Thus, we have demonstrated the influence of absorption linewidth enhancement
in semiconductor on the ultrashort pulse characteristics in solid-state
lasers. The linewidth enhancement introduces a negative feedback that
stabilizes the pulse. The region of the pulse stability is much wider in
this case than in the case of the soliton stabilization. At low pump powers
and low negative GVD a dramatic growth of the pulse duration is observed.
Bistable operation for large pump powers is possible. The advantage of
mode-locking mechanism described above is the operation without
Kerr-lensing, which is very attractive for diode-pumped cavity alignment
insensitive system with large mode cross section.

\section{Ultrashort pulse generation in the presence of the quadratic ac
Stark effect}

An absorption linewidth enhancement is not a single nonlinear mechanism in
semiconductors that may contribute to mode-locking. Estimations show that
the Stark shift of the excitonic resonance in the presence of strong laser
field can be a strong pulse-shaping factor, too [5, 12]. Furthermore, an
experimental utilization of the Stark effect due to external modulation was
successfully used for active mode-locking of diode-pumped Nd: YAG laser [13].

\subsection{Model}

Quadratic ac Stark shift $\Delta \omega $ that is due to the influence of
nonresonant transitions on excitonic resonance is proportional to the
polarizability difference between ground and excited states $\Delta \alpha $
[14]: $\Delta \omega =|\Delta \alpha |\times |E|^{2}/h$, where $E$ is the
field strength. The typical values of $|\Delta \alpha |$ for semiconductors
are of order $10^{-19}\div $ $10^{-21}$ $cm^{3}$ [15], that corresponds to
the Stark shift coefficient $\zeta =8\pi |\Delta \alpha |/(nch)=(10^{5}\div
10^{3})/n\;cm^{2}J^{-1}$, where $n$ is the index of refractivity. Neglecting
the population of higher energy levels of semiconductor and under weak
exciton-exciton, exciton-phonon bound approximation, one can describe the
interaction between laser field and quantum-confined absorber by the
generalized two-level model [14]. The evolution of the off-diagonal element
of the density matrix $\Pi $ and the population difference between ground
and excited states $\Xi $ obey the differential equation set:

\begin{eqnarray}
\frac{d\Pi }{dt}+[\frac{1}{t_{a}}-i(\omega _{l}-\omega _{a}-\Delta \omega
)]\Pi  &=&\frac{i}{h}\wp \Xi , \\
\frac{d\Xi }{dt}+\frac{\Xi -\Xi _{0}}{T} &=&-\frac{4}{h}Im(\Pi \wp ^{\ast }),
\nonumber
\end{eqnarray}

where $t_{a}$ is the inverse bandwidth of the absorption line, $\omega _{a}$
is the resonance frequency, $\omega _{l}$ is the field carrier frequency, $%
\wp $ is the matrix element of the interaction, $\Xi _{0}$ is the
equilibrium population difference. The Stark shift which is possible only in
generalized two-level model, in quasi-monochromatic approximation, i. e.
under neglecting the cross-modulation between different pulse spectral
components, is proportional to $\varsigma \left| a(t)\right| ^{2}$.

In the noncoherent approach, the evolution of the complex field $a(t)$ in
the laser system containing the gain medium, saturable absorber, frequency
filter and dispersive element obeys nonlinear operator's equation: $%
a^{k+1}(t)=\tilde{\Gamma}\tilde{G}\tilde{D}a^{k}(t),$ where $k$ is the round
trip number, $t$ is the local time. The Lorenzian gain band action is
described by $\tilde{G}=\exp (\frac{\alpha L_{g}}{1+L_{g}t_{g}\frac{\partial 
}{\partial t}})$, $L_{g}=\frac{1}{1+i(\omega _{l}-\omega _{g})t_{g}}$ where $%
\omega _{g}$ is the gain resonance, $t_{g}$ is the inverse gain bandwidth
(for Cr: forsterite chosen for our calculations, $t_{g}$ $=20\;fs$), $\alpha 
$ is the saturated gain. $\tilde{D}=\exp (id\frac{\partial ^{2}}{\partial
t^{2}})$ is the second-order GVD group velocity dispersion operator, where $d
$ is the dispersion amount introduced by the prism pair. From eq. (11) one
derives an operator accounting for the effects of saturable absorption and
nonlinear contribution due to Stark-effect\newline
$\Gamma =\exp {\frac{-\gamma L_{a}\exp
[-Re(L_{a})\int\limits_{t_{0}}^{t}\left| a(t\prime )\right| ^{2}\exp (-\frac{%
t-t\prime }{T})dt\prime /U_{a}]}{1+L_{a}t_{a}\frac{\partial }{\partial t}}-}%
\mathit{l}$ , where\newline
$L_{a}=\frac{1}{1+i[\omega _{l}-(\omega _{a}+\mu \left| a(t)\right|
^{2})]t_{a}}$ , $U_{a}$ is the saturation energy, $\gamma $ is the saturated
loss at the moment $t_{0}$ corresponding to the pulse peak, $l$ is the
linear loss. The integral in $\Gamma $ accounts for the ordinary slow
saturable absorption with recovery time $T$. Here we used the normalization
as in previous section, so that for PbS -- based modulator with $%
U_{a}=390\;\mu J/cm^{2}$ [16] $\mu $ $=\zeta U_{a}=13$.

An expansion of laser equation into series on $t$, local field energy and
intensity, provided that the pulse duration is much shorter than $T$, gives
the laser dynamical equation similar to the generalized Landau-Ginzburg
equation:

\begin{eqnarray}
\frac{\partial a(k,t)}{\partial k} &=&[c_{1}+ic_{2}\frac{\partial }{\partial
t}+(c_{3}+ic_{4})\frac{\partial ^{2}}{\partial t^{2}}+ \\
&&(c_{5}+ic_{6})\left| a(k,t)\right| ^{2}+(c_{7}+ic_{8})\epsilon +  \nonumber
\\
&&(c_{9}+ic_{10})\frac{\epsilon ^{2}}{2}+(c_{11}+ic_{12})\epsilon \frac{%
\partial }{\partial t}]a(k,t),  \nonumber
\end{eqnarray}

where $\epsilon =\int\limits_{t_{0}}^{t}\left| a(k,t^{\prime })\right|
^{2}dt^{\prime }$, $c_{1}=\alpha J_{g}-\gamma J_{a}-l$, $c_{2}=2\alpha
\Omega J_{g}^{2}-2\gamma \omega \vartheta ^{2}J_{a}^{2}$, $c_{3}=(1-3\Omega
^{2})J_{g}^{3}-\gamma (1-3\omega ^{2}\vartheta ^{2})J_{a}^{3}\vartheta ^{2}$%
, $c_{4}=\alpha (\Omega ^{3}-3\Omega )J_{g}^{3}-\gamma (\omega ^{3}\vartheta
^{3}-3\omega \vartheta )J_{a}^{3}\vartheta ^{2}+d$, $c_{5}=-2\gamma \omega
\vartheta \mu H_{a}$, $c_{6}=-\gamma \mu (1-\omega ^{2}\vartheta ^{2})H_{a}$%
, $c_{7}=\gamma J_{2}^{2}$, $c_{8}=-\gamma \omega \vartheta J_{a}^{2}$, $%
c_{9}=-\gamma J_{a}^{3}$, $c_{10}=\gamma \omega \vartheta J_{a}^{3}$, $%
c_{11}=-\gamma (1-\vartheta ^{2}\omega ^{2})J_{a}^{3}\vartheta $, $%
c_{12}=2\omega \gamma \vartheta ^{2}J_{a}^{3}$, $J_{a}=\frac{1}{1+\omega
^{2}\vartheta ^{2}}$, $J_{g}=\frac{1}{1+\Omega ^{2}}$, $H_{a}=\frac{1}{%
(1-\omega ^{2}\vartheta ^{2})^{2}+4\omega ^{2}\vartheta ^{2}}$, $\omega
=\omega _{l}-\omega _{a}$, $\Omega =\omega _{l}-\omega _{g}$, $\vartheta
=t_{a}/t_{g}$. We normalized all times to $t_{g}$, frequencies to $t_{g}^{-1}
$, dispersion coefficient to $t_{g}^{2}$, the intensity to $\frac{U_{a}}{%
t_{g}}$.

\subsection{Stark induced mode locking}

As one can see, Eq. (12) contains the term $-2\mu \gamma H_{a}\omega
\vartheta \left| a(k,t)\right| ^{2}a(k,t)$ , which is responsible for the
fast saturable absorption at the below band-gap excitation with $\omega <0$
(i. e. red shift of the pulse carrier frequency from the excitonic
resonance). The shift $\omega $ is due to detuning between the gain and loss
resonances ($\omega _{g}-\omega _{a}<0$). The fast saturable absorber action
is caused by the induced ``pushing out'' of the excitonic resonance from the
red-shifted pulse spectrum due to power-dependent blue Stark shift. The
corresponding saturation intensity is $I_{s}=[-2\mu \gamma H_{a}t_{a}(\omega
_{l}-\omega _{a})]^{-1}$. Our calculations showed, that the saturation
intensity is close to the level, which is typical for Kerr-lens mode-locked
lasers and is high enough to provide self-starting.

Eq. (12) has quasi-solution solution $a(k,t)=a_{0}sech^{1+i\psi }[(t-k\delta
)/t_{p}]e^{i\phi z}$, where the pulse parameters have the same meaning as in
section 2.2. The duration and frequency shift for chirp-free solution and
necessary GVD are presented in Fig. 5. It is seen, that the minimal pulse
durations (close to the limit defined by $t_{g}$) are provided by $(\omega
_{a}-\omega _{g})t_{g}=0.3\div 0.7$. The growth of the initial unsaturated
loss $\Gamma $ shortens the pulse duration (curve 2 in compare with 1). The
minimum of pulse duration in $\omega _{a}-\omega _{g}$ approximately
coincides with the minimum of $I_{s}$. Using of smaller $U_{a}$ increases
the duration (curve 3) because the absorber operates in the regime of strong
saturation. A stronger saturation results in an additional blue shift of the
pulse with respect to the gain resonance (curve 3 of Fig. 5, b) (for the
explanation of the mechanism of this shift see [17]), which does not favor
the Stark induced mode locking. As is seen from eq. (12), ac Stark effect
contributes both to SPM-term $c_{6}$ and GVD-term $c_{4}$, which modifies
the amount of GVD from prism pair necessary for chirp compensation (Fig. 5,
c).

Using the absorber with narrower band in compare with gain band (curve 4)
reduces the region of $\omega _{g}-\omega _{a}$, where chirp-free pulses
exist because the pulse spectrum and excitonic line do not overlap. However,
in this case the generation of the chirped pulses with duration close to $%
t_{g}$ is possible (Fig. 6). The larger $\tau $ requires positive dispersion 
$d$ for pulse existence (curve 2 in compare to 1), which is explained by the
change of the sign of self-phase modulation coefficient in $c_{6}$. It
should be noted, however, that for the case of large $\tau $ it is be
necessary to account for the coherent nature of pulse-semiconductor
interaction (here such effects as self-induced transparency are possible
[18]), which indeed transforms mode-locking dynamics essentially, but falls
out the frame of the present model.

The predicted dependence of the pulse parameters on the frequency shift from
excitonic resonance is confirmed by experimental results presented in [12].
Here, in particular, there is no ultrashort pulse generation at no shift
from excitonic resonance, a relatively small Stokes shift of the laser
frequency from the excitonic resonance caused fs-pulse operation, while the
increase of the Stokes shift brings the pulse duration into picosecond
region. Our conclusions that the strong saturation of semiconductor
absorption does not favor the Stark-induced amplitude modulation also
corroborates with experimental results [5].

In order to perform the stability analysis we calculated the net-gain $%
\Sigma $ behind the pulse tale. Evidently, the pulse would be unstable, if $%
\Sigma >$0 for some frequency $\omega _{n}$ from the noise spectral region.
Fig. 7 shows $\Sigma $ for five selected noise frequencies ($\Delta =\omega
_{n}-\omega _{l}$). It is seen that the decrease of $\Gamma $ (Fig.7, a in
compare with b), decrease of $U_{a}$ (Fig. 7, c in compare with Fig. 7, b)
or growth of $\vartheta $ (Fig. 7, d) destabilizes the chirp-free pulse, but
the suitable detuning between absorber and gain resonances $\omega
_{a}-\omega _{g}$ allows for stable ultra-short pulse generation with
duration much shorter than the absorber recovery time.

So, the quadratic Stark effect in semiconductor absorbers causes a stable
ultrashort pulse generation in case of relatively small absorption
saturation and below band-gap excitation. The generation of chirp-free
pulses with the duration close to the theoretical minimum is possible.

Next section is devoted to the case of the strong absorber saturation which
is the typical situation for real antiresonant Fabry-Perot saturable
absorber mirrors [4] and requires some modifications to be introduced in our
model.

\section{Ultrashort pulse generation in the presence of linear ac Stark
effect}

Here we shall demonstrate that the linear Stark effect at near-resonance
interaction with semiconductor absorber can efficiently reduce the threshold
of ultrashort pulse generation with the durations close to shortest
possible. We present a theory for the formation of ``weak-nonlinear''
soliton with GVD balanced by SPM which is proportional rather the amplitude
than the intensity of the field.

As is known [19], an optical Stark effect in semiconductors with reduced
dimension is described in frame of ``dressed-exciton'' model [19, 20]. In
this model the shift of excitonic resonance originates from the mixing of
``field - matter'' states. At below/above band-gap excitation a blue/red
shift is observed (in the latter case the driving field falls into the
absorption band of semiconductor which complicates an experimental
observation of red shift), a precise coincidence between driving frequency
and excitation resonance causes the splitting of resonance. The magnitude of
the Stark shift is $\sqrt{\omega ^{2}+\varsigma \left| a\right| ^{2}}$ ,
where $\omega $ is the mismatch from resonance, $a$ is the field, $\varsigma 
$ is the Stark shift coefficient (i. g. for GaAs/AlGaAs quantum well $%
\varsigma =4\times 10^{16}\;Hz\cdot cm^{2}/J$ [21]). As was shown in [19,
13], the Stark shift produces almost noninertial amplitude modulation which
may be used in ultrafast optical modulators.

\subsection{Model}

To describe the Stark shift of excitonic resonance at near-resonance
interaction ($\omega \longrightarrow 0$) with semiconductor absorber we
adopted the following approximations: i) an excitonic bond has a Lorentzian
profile, ii) the exciton-exciton bond is weak and therefore neglected, and
iii) the field is quasi-monochromatic. For saturation flux of absorber we
used $U_{a}=100\;\mu J/cm^{2}$, the relaxation time was much longer than the
pulse duration. With the above assumptions, an operator for the interaction
of the field with excitonic resonance is (see section 3.1):

\vspace{1pt}%
\begin{equation}
\Gamma =-\frac{L_{a}\exp [-ReL_{a}\int\limits_{-\infty }^{t}\left| a(t\prime
)\right| ^{2}dt\prime /U_{a}]}{1+L_{a}t_{a}\frac{\partial }{\partial t}}
\end{equation}

where $L_{a}=\frac{1}{1+i\kappa t_{a}\left| a(t\prime )\right| }$ is the
saturable loss, $\kappa =\sqrt{\varsigma }$ , $t_{a}$ is the inverse
absorption bandwidth, $t$ is the local time.

Real semiconductor saturable absorber mirrors (SESAM's) operate in the
condition of strong saturation [22], and one may therefore neglect the
dynamical saturation and introduce a saturated by the full pulse energy loss 
$\gamma $:\newline
$\gamma =\frac{\Gamma }{1+ReL_{a}\int\limits_{-\infty }^{\infty }\left|
a(t\prime )\right| ^{2}dt\prime /U_{a}}$.

Thus, the basic pulse shaping mechanism here is that of soliton formation
with the SPM balanced by the GVD, but the SPM now has two components, cubic
in field (Kerr-nonlinearity in active medium) and quadratic in field
(Stark-shift induced nonlinearity in semiconductor), the latter essentially
transforming the nature and parameters of the pulse.

Expansion of eq. (11) into the power series in local time $t$ and field $a$
and retaining the terms up to the 2nd order yields:

\vspace{1pt}%
\begin{eqnarray}
\frac{\partial a(t,k)}{\partial k} &=&\left[ \alpha -\gamma -l+i\varphi %
\right] a(t,k)+ \\
&&\left[ (\alpha -\gamma +id)\frac{\partial ^{2}}{\partial t^{2}}\right]
a(t,k)\mp  \nonumber \\
&&\left[ i\gamma \left| a\right| +(\gamma -ip)\left| a\right| ^{2}\right]
a(t,k),  \nonumber
\end{eqnarray}

where $k$ is round-trip number, $\alpha $ is the saturated gain, $l$ is the
linear loss, $\varphi $ is the phase delay, $d$ is the GVD amount, $p$ is
coefficient of Kerr-nonlinearity $\beta =\frac{2\pi n_{2}z}{\lambda n}$
normalized to $t_{a}^{2}\varsigma $ ($n_{2}$ and $n$ are nonlinear and
linear refraction indices an active medium, $z$ is the length of active
medium, $\lambda $ is the generation wavelength), the signs ``$-$`` and ``$+$%
'' correspond to below and above band-gap excitation. The saturated gain $%
\alpha $ is calculated through the gain at full inversion as\newline
$\alpha =\alpha _{m}\frac{1-\exp (-U)}{1-\exp (-U-\sigma
\int\limits_{-\infty }^{\infty }\left| a\right| ^{2}dt)}\exp (-\tau
\int\limits_{-\infty }^{\infty }\left| a\right| ^{2}dt/2)$, where $U$ is the
pump photon flux, normalized to cavity period $T_{cav}$ and absorption
cross-section $\sigma _{13}$, $\tau ^{-1}=U_{g}\varsigma t_{a}$ is
normalized energy flux of the gain saturation. In eq. (14) the time and
intensity are normalized to $t_{a}$ and and $t_{a}^{2}\varsigma $ ,
respectively. We assumed also that the gain line has the Lorentzian profile
and inverse bandwidth is equal to that of semiconductor absorber $t_{a}$.
For Cr: forsterite laser ($t_{a}=20\;fs$) where an ultrashort pulse
generation using the bleaching of excitation resonance in PbS quantum dots
has been reported [12] with our normalizations and 3-mm long crystal the
parameters $\tau $ and $p$ are 0.01 and 0.1, respectively.

It is seen from eq. (14), that the Stark shift of excitonic resonance which
may be understand as pulse-induced `` pushing-out'' of absorption band form
the pulse spectral profile introduces nonlinear (quadratic) phase modulation
as well as fast saturable loss proportional to intensity. The sign of SPM in
absorber coincides with that produced by Kerr-nonlinearity in active medium
for the above band-gap excitation and is opposite for the below band-gap
excitation.

In the general solution for eq. (14) is unknown, so we shall consider two
limiting cases of eq. (14): i) weak generation field and ii) strong
generation field. In the first case one may neglect the cubic in field term:

\vspace{1pt}%
\begin{eqnarray}
\frac{\partial a}{\partial k} &=&\left[ \alpha -\gamma -l+i\varphi \right] a+
\\
&&\left[ (\alpha -\gamma +id)\frac{\partial ^{2}}{\partial t^{2}}\right] a\mp
\nonumber \\
&&i\gamma \left| a\right| a.  \nonumber
\end{eqnarray}

In the second case one may neglect the quadratic in field term:

\vspace{1pt}%
\begin{eqnarray}
\frac{\partial a}{\partial k} &=&\left[ \alpha -\gamma -l+i\varphi \right] a+
\\
&&\left[ (\alpha -\gamma +id)\frac{\partial ^{2}}{\partial t^{2}}\right] a+ 
\nonumber \\
&&(\gamma -ip)\left| a\right| ^{2}a  \nonumber
\end{eqnarray}

\subsection{Non-Schrödinger soliton in a solid-state laser with
semiconductor saturable absorber}

Eq. (15), which is analogues to the Fisher-equation [23], has a
quasi-soliton solution in the form:

\vspace{1pt}%
\begin{equation}
a(t)=a_{0}sech^{2+i\psi }(t/t_{p}),
\end{equation}

where the pulse amplitude $a_{0}$, the chirp $\psi $ and the duration $t_{p}$
relate as follows:

\vspace{1pt}%
\begin{eqnarray}
\psi &=&\frac{5d\pm \sqrt{25d^{2}+24(\alpha -\gamma )^{2}}}{2(\gamma -\alpha
)}, \\
t_{p} &=&\sqrt{\frac{2(\alpha -\gamma )-d\psi }{\alpha -\gamma -l}}, 
\nonumber \\
a_{0} &=&\frac{5\psi \left[ (\alpha -\gamma )^{2}+d^{2}\right] }{\gamma
t_{p}^{2}(\gamma -\alpha )},  \nonumber
\end{eqnarray}

The signs ``$-$'' and ``$+$'' have the same meaning as in eq. (14).

Eq. (16) is the equation for laser Schrödinger soliton generated in the
system with the gain, loss, Kerr-nonlinearity and GVD [24]. The solution for
eq. (16) is

\vspace{1pt}%
\begin{equation}
a(t)=a_{0}sech^{1+i\psi }(t/t_{p}),
\end{equation}

with

\vspace{1pt}%
\begin{eqnarray}
a_{0} &=&\sqrt{\frac{(\gamma -\alpha )(\psi ^{2}-2)-3d\psi }{\gamma t_{p}^{2}%
}}, \\
t_{p} &=&\sqrt{\frac{(\gamma -\alpha )(\psi ^{2}-1)-2d\psi }{\gamma -\alpha
+l}},  \nonumber \\
\psi &=&\{3[dp+\gamma (\gamma -\alpha )]+  \nonumber \\
&&\sqrt{9[dp+\gamma (\gamma -\alpha )]^{2}+8[p(\gamma -\alpha )-\gamma d]^{2}%
}\}/  \nonumber \\
&&2[p(\gamma -\alpha )-\gamma d].  \nonumber
\end{eqnarray}

Pulse duration and chirp vs GVD are presented in Figs. 8 and 9. Solid curves
1 and 2 correspond to the case of weak field and below and above band-gap
excitation, respectively. Dashed lines correspond to the case of strong
field.

From Fig. 9 (solid line 1 and dashed line) one may see that at below
band-gap excitation the signs of the chirp produced by Stark-shift in
semiconductor and Kerr-nonlinearity in active medium coincide. In the first
limiting case this reduces the pulse duration in the region of negative GVD.
Characteristic features of solution of the type (17) are that i) the pulse
duration is close to the minimal possible $t_{a}$, and the necessary pump
rates are much lower than for pure quasi-Schrödinger soliton, that ii) full
chirp compensation due to GVD is impossible and that iii) the GVD range
where the solution exists is much narrower than that one for quasi-Schrö%
dinger soliton.

Solution (17) is transformed essentially in the case of the above band-gap
excitation (curves 2 in Figs. 8 and 9): the chirp changes its sign according
to ``defocusing'' action of SPM in semiconductor and reaches its minimum
(along with the pulse duration) at some positive GVD; the GVD range where
the solution exists narrows as compare to the previous case.

To check the correctness of the approximation of the general equation (14)
by eq. (15) for weak-field limit we investigated the behavior of the
parameters of quasi-soliton (17) in eq. (15) perturbed by cubic term.
Substitution of (17) into (14) and expansion into the time series yields an
algebraic equations set:

\vspace{1pt}%
\begin{eqnarray}
\lefteqn{2(\gamma -\alpha )+d\psi -(\gamma -\alpha +l)t_{p}^{2}+\gamma
a_{0}^{2}t_{p}^{2} =0,} \\
&&-6(\gamma -\alpha )-5d\psi +(\gamma -\alpha )\psi ^{2}-2\gamma
a_{0}^{2}t_{p}^{2} =0,  \nonumber \\
&&-12d+4(\gamma -\alpha )\psi -3d\psi ^{2}+(\gamma -\alpha )\psi
^{3}-2a_{0}t_{p}^{2}\gamma -2a_{0}^{2}t_{p}^{2}(2p+\psi \gamma ) =0. 
\nonumber
\end{eqnarray}

The pulse duration and the chirp obtained from these equations are presented
by dotted lines in Figs. 8 and 9. It is seen, that for the weak-field
approximation the cubic term is really just a perturbation and changes the
solution only slightly. One can see also, that when the signs of SPM in
active medium and in semiconductor coincide (above band-gap excitation), the
decrease of pulse duration is observed, while for the opposite signs (below
band-gap excitation) the duration increases.

Analysis of the pulse stability against perturbation of its parameters
(amplitude, duration and chirp) and energy was performed is follows.
Substituting expression (17) into eq. (14) and expanding it into the time
series we arrive to the equations set for the pulse parameters evolution. As
before (Sec. 2.2) the condition for the pulse stability against small
parameters perturbations is the real part of Jacobean of the system to be
nonpositive. After multiplying eq. (14) by complex-conjugate field, summing
up with complex -conjugate expression and integrating over full time we get
the conservation law for the pulse energy from which the condition for small
energy perturbation evolution follows.

All solutions presented in Figs. 8 and 9 satisfy the above stability
criterion. As analysis shows, the main destabilizing factor is $-d\psi $ ,
however at small negative GVD it may be dominated by stabilizing factors
(gain, loss, spectral filtering and saturable loss) thus making the stable
ultrashort pulse generation possible.

In order to investigate the pulse characteristics in the intermediate region
of field intensities we sought for the approximate solution of eq. (14) in
the form

\vspace{1pt}%
\begin{equation}
a(t)=a_{0}\exp (-(t/t_{p})^{2}+i\psi t^{2}),
\end{equation}

\vspace{1pt}

The pulse parameters are calculated then through the substitution of
expression (22) into eq. (14), the expansion into the time series and
solving the algebraic equations set for the unknowns $a_{0}$, $t_{p}$, $\psi 
$. The first two relations read:

\vspace{1pt}%
\begin{eqnarray}
a_{0} &=&\{4[d\gamma +(\alpha -\gamma )p+(l+\gamma -\alpha )t_{p}^{2}p/2+ \\
&&2\psi t_{p}^{2}(\gamma (\alpha -\gamma )-dp)-d\gamma \psi ^{2}t_{p}^{4}]\}/
\nonumber \\
&&\left[ -\gamma ^{2}t_{p}^{2}\right] ,  \nonumber \\
t_{p} &=&\sqrt{\frac{\alpha -\gamma -l-2d\psi }{2\psi ^{2}(\gamma -\alpha )}}%
.  \nonumber
\end{eqnarray}

The duration of the stable pulse vs pump power is presented in Fig. 10,
where solid curves 1 and 2 correspond the below and above band-gap
excitation, respectively, dashed line is the pulse duration for the case
with no Stark-effect. It is seen, that for the case of the below band-gap
excitation the contribution of quadratic SPM in semiconductor essentially
reduces the pulse duration as compare to the case with no Stark-effect, so
that one can achieve a short pulse generation (with duration of 100$\div $%
200 fs) even at low pump powers (curve 1 and dashed curve). For the case of
the above band-gap excitation an abrupt switch from ps- to fs-regime is
observed at the pump-power for which the cubic nonlinearity begins to
dominate over the quadratic one (compare this behavior with that described
in [24]).

Thus, we demonstrated that the linear Stark-shift of excitonic resonance at
near resonance interaction of ultrashort pulses with semiconductor saturable
absorber in cw solid-state laser transforms the pulse characteristics
essentially: due to formation of `` weak-nonlinear'' optical soliton. At low
pump power and the below band-gap excitation a remarkable decrease of the
pulse duration takes place, while at the above band-gap excitation a switch
from ps- to fs-regime is observed.

\section{Conclusions}

Nonlinear properties of semiconductor saturable absorbers, such as linewidth
enhancement, quadratic and linear ac Stark effect contribute essentially to
the mode locking in cw solid-state lasers. The linewidth enhancement due to
carrier density induced slow nonlinear refraction produces a negative
feedback, which stabilizes ultrashort pulse against automodulational
instabilities and expands the region of the pulse existence. Besides, this
nonlinear factor can produce multistable lasing and switching to regime of
the picosecond pulse generation.

At relatively small absorption saturation, the quadratic ac Stark shift of
absorption line produces a fast saturable absorber action. The saturation
intensity of this fast saturation is comparable to that of typical Kerr-lens
mode locked lasers. For the case of strong absorption saturation the linear
ac Stark effect produces a ``weak-nonlinear'' soliton at relatively small
pulse intensities, that reduces the threshold of the ultrashort pulse
formation.

The main advantages of lasing regimes described above are the operation
without Kerr-lensing, which is very attractive for diode-pumped cavity
alignment insensitive system with large mode cross section, the
self-starting ability and the relatively small thresholds for fs-pulse
generation.

\section{References}

[1] I. D. Jung, F. X. Kärtner, N. Matuschek, D. H. Sutter, F. Morier-Genoud,
V. Scheuer, M. Tillich, T. Tschudi and U. Keller, Appl. Phys. B 65 (1997)
137.

[2] U. Morgner, S. H. Cho, J. Fini, F. X. Kärtner, H. A. Haus, J. G.
Fujimoto, E. P. Ippen, V. Scheuer, M. Tilsch, T. Tschudi, Tech. Dig.
Conference Advanced Solid-State Lasers, 1999, paper TuA3, p. 198.

[3] P. T. Guerreiro, S. Ten, E. Slobodchikov, Y. M. Kim, J. C. Woo, N.
Peyghambarian, Optics Commun. 136 (1997) 27.

[4] F. X. Kärtner, I. D. Jung and U. Keller, IEEE J. Selected Topics in
Quant. Electr. 2 (1996) 540.

[5] W. H. Knox, D. S. Chemla, D. A. B. Miller, J. B. Stark, S. Schmitt-Rink,
Phys. Rev. Lett. 62 (1989) 1189

[6] H. Haug, S. W. Koch, Quantum theory of the optical and electronic
properties of semiconductors (World Scientific, 1994).

[7] C. H. Henry, IEEE J. Quant. Electr. QE-18 (1982) 259.

[8] M. Osinski, J. Buus, IEEE J. Quant. Electr. QE-23 (1987) 9.

[9] H. A. Haus, IEEE J. Quant. Electr. QE-11 (1975) 736.

[10] S. H. Namiki, E. P. Ippen, H. Haus, C. X. Yu, J. Opt. Soc. Am. B14
(1997) 2099.

[11] J. P. Gordon, J. Opt. Soc. Am. B9 (1992) 91.

[12] S. Tsuda, W. H. Knox, S. T. Cundiff, W. Y. Jan, and J. E. Cunningham,
IEEE J. Selected Topics in Quant. Electr. 2 (1996) 454.

[13] L. R. Brovelli, M. Lanker, U. Keller, K. W. Goossen, J. A. Walker, and
J.E. Cunningham, Electron. Lett. 31 (1995) 381.

[14] V. S. Butilkin, A. E. Kaplan, Yu. G. Chronopulo, E. P. Jakubovich,
Rezonansnie vzaimodejstvija sveta s veschestvom (Moscow: Nauka,1977) p 38.

[15] P. G. Elyseev, A. P. Bogatov, Trudi FIAN 166 (1986) 15.

[16] P. T. Guirreiro, S. Ten, N. F. Borrelli, J. Butty, G. E. Jabbour, N.
Peyghambarian, Appl. Phys. Lett. 71 (1997) 1595.

[17] V. L. Kalashnikov, V. P. Kalosha, I. G. Poloyko, and V. P. Mikhailov,
J. Opt. Soc. Am. B14 (1997) 2112.

[18] V. L. Kalashnikov, D. O. Krimer, I. G. Poloyko, arXiv: physics/ 0009006.

[19] A. Mysyrowicz, D. Hulin, A. Antonetti, A. Migus, W. Masselink, H.
Morkoc, Phys. Rev. Lett. 56 (1986) 2748.

[20] C. Cohen-Tannoudji, S. Reynaud, J. Phys. B: Atom. Molec. Phys. 10
(1977) 345.

[21] A. von Lehmen, D. S. Chemla, J. E. Zucker, J. P. Heritage, Opt. Lett.
11 (1986) 609.

[22] U. Keller, K. J. Weingarten, F. X. Kärtner, D. Kopf, B. Braun, I. D.
Jung, R. Fluck, C. Hönninger, N. Matuschek, and J. aus der Au, IEEE J. of
Selected Topics in Quant. Electron., 2 (1996) 435.

[23] M. Ablowitz, H. Segur, Solitons and inverse scattering problem,
(Philadelphia 1987).

[24] V. L. Kalashnikov, D. O. Krimer, I. G. Poloyko, J. Opt. Soc. Am., B 17
(2000) 519 (see also arXiv: physics/ 0009009).

\newpage

\section{Figure captions}

Fig. 1. Chirp $\psi $ versus pump power for different linewidth enhancement
factors: $\chi =$0\ (1),\ -0.005\ (2),\ -0.05\ (3),\ -2.5\ (4). Every
parameters set has two solutions. Stable solutions are plotted by solid
lines. GVD coefficient is -360 fs$^{2}$, $\Gamma =$0.05,$\;l=$0.05,$\;a_{m}=$%
1.5,$\;p=$3.

Fig. 2. Frequency shift $\omega $ versus pump power for different linewidth
enhancement factors and GVD coefficients: $\chi =$0\ (1),\ -0.005\ (2),\
-0.05\ (3),\ -2.5\ (4,\ 5); $d=$-360\ fs$^{2}$\ (1 - 4),\ -90\ fs$^{2}$\
(5). Other parameters are as in Fig. 2. Stable solutions are plotted by
solid lines. To better illustrate the behavior of the curves the plot is
divided into parts \textit{a} and \textit{b}.

Fig. 3. Pulse duration $t_{p}$ versus pump power for different linewidth
enhancement factors and GVD coefficients: $\chi =$0\ (1),\ -0.005\ (2),\
-0.05\ (3),\ -2.5\ (4,\ 5); $d=$-360\ fs$^{2}$\ (1 - 4),\ -90\ fs$^{2}$\
(5). Other parameters are as in Fig. 2. Stable solutions are plotted by
solid lines.

Fig. 4. Regions of pulse existence. (\textit{A}) -- region of instability, (%
\textit{B}) - automodulational stability, (\textit{C}) - stability against
noise, (\textit{D}) -- automodulational and noise stability on the plane
(GVD -- pump power). $\chi =$0\ (\textit{a}),\ -2.5\ (\textit{b}).

Fig. 5. Duration $t_{p}$ (\textit{a}), GVD amount $d$ (\textit{b}) and
normalized frequency mismatch from excitonic resonance $\Omega t_{g}$ (%
\textit{c}) for chirp-free solution versus normalized mismatch between gain
and absorption resonances $(\omega _{a}-\omega _{g})\times t_{g}$. $\chi =$%
13\ (1,\ 3,\ 4),\ 8\ (2), $\Gamma =$0.05\ (1,\ 2),\ 0.1\ (3,\ 4), \ $%
\vartheta =$1\ (1-3),\ 3\ (4), $\alpha-r=$0.01.

Fig. 6. Chirp $\psi $ (\textit{a}) and duration of the pulse $t_{p}$ (%
\textit{b}) versus GVD amount $d$. $\vartheta $ $=$1\ (1),\ 30\ (2), $%
(\omega _{a}-\omega _{g})\times t_{g}=$0.5\ (1),\ 0.2\ (2); $\chi =$13, $%
\Gamma =$0.1, $\alpha-r=$0.001.

Fig. 7. Net-gain $\Sigma $ behind the pulse tail for five selected noise
frequencies $(\omega _{n}-\omega _{l})\times t_{g}=$1 (curve1),\ 0.5\ (2),
0\ (3),\ -0.5\ (4),\ -1\ (5). $\chi =$13\ (\textit{a},\ \textit{c},\ \textit{%
d}),\ 8\ (b), $\Gamma =$0.05\ (\textit{a},\ \textit{b}),\ 0.1\ (\textit{c},\ 
\textit{d}), $\vartheta =$1\ (\textit{a},\ \textit{b},\ \textit{c}),\ 3\ (%
\textit{d}).

Fig. 8. Duration of the stable pulse $t_{p}$ versus GVD $d$. Curve 1 - pulse
(17) at below band-gap excitation, curve 1 - pulse (17) at above band-gap
excitation, dashed curve - pulse (19), dotted curves - pulse (17) as
solution for eq. (15) perturbed by cubic nonlinearity. Pump power is 310 mW
(solid and dotted curves) and 9.7 W (dashed curve), pump beam radius is 50 $%
\mu $m, $\alpha _{m}=$1, $\Gamma $ $=\;$0.1, $l=$ 0.05.

Fig. 9. Chirp $\psi $ of the stable pulse versus GVD $d$. Curve 1 - pulse
(17) at below band-gap excitation, curve 1 - pulse (17) at above band-gap
excitation, dashed curve - pulse (19) , dotted curves - pulse (17) as
solution for eq. (15) perturbed by cubic nonlinearity. Other parameters are as
in Fig. 8.

Fig. 10. Duration of the stable pulse (22) $t_{p}$ versus pump power. Curve
1 - below band-gap excitation, curve 1 - above band-gap excitation, dotted
curve - the case with no Stark-effect in semiconductor, $d=$ -15.6 fs$^{2}$
and other parameters as in Fig. 8.

\begin{figure}
	\begin{center}
		\includegraphics{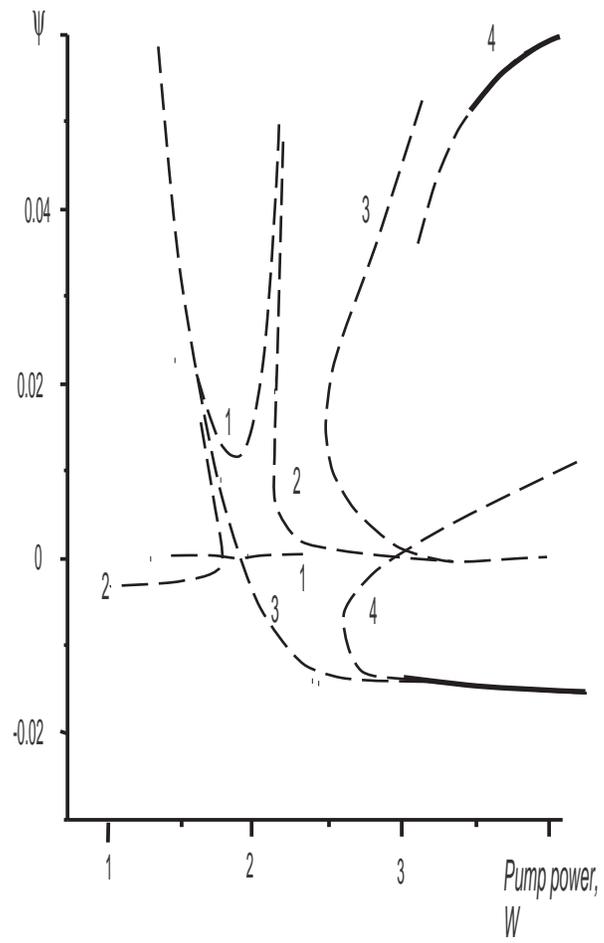}
	\end{center}

	\caption{Chirp $\psi $ versus pump power}
\end{figure}

\begin{figure}
	\begin{center}
		\includegraphics{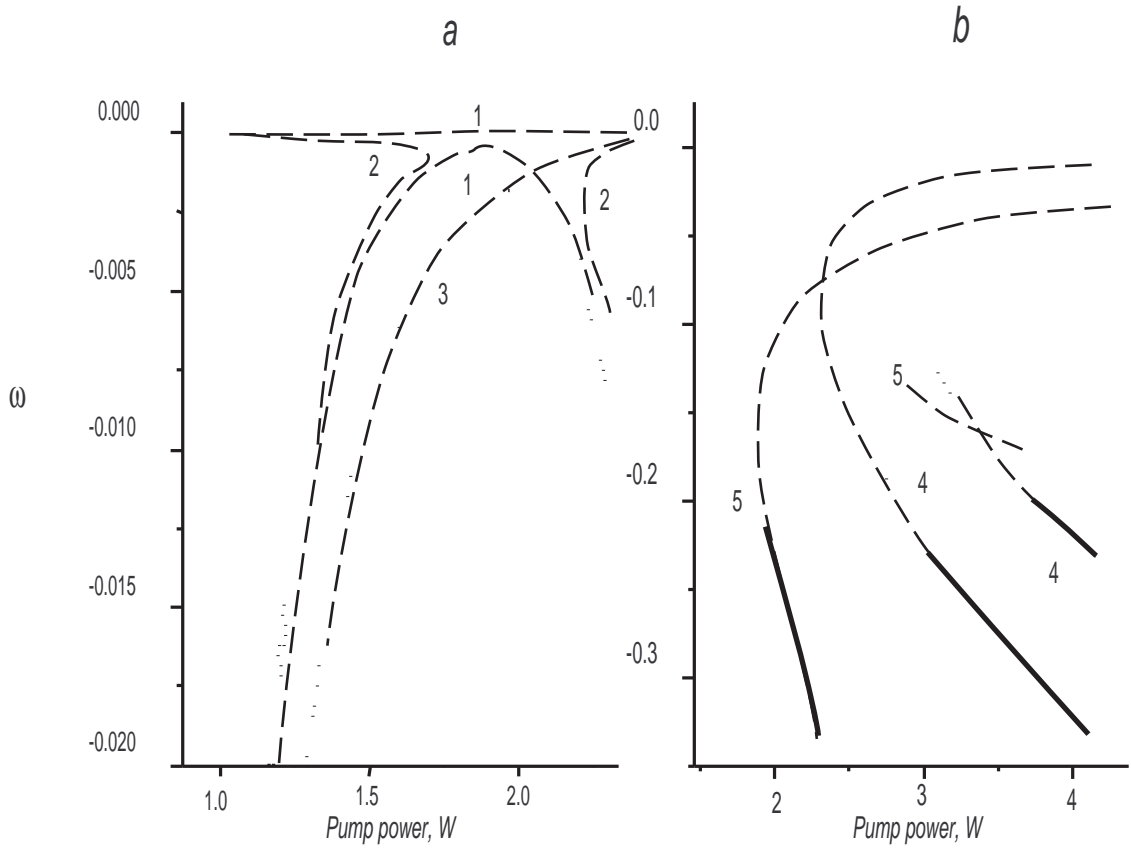}
	\end{center}

	\caption{Frequency shift $\omega $ versus pump power}
\end{figure}

\begin{figure}
	\begin{center}
		\includegraphics{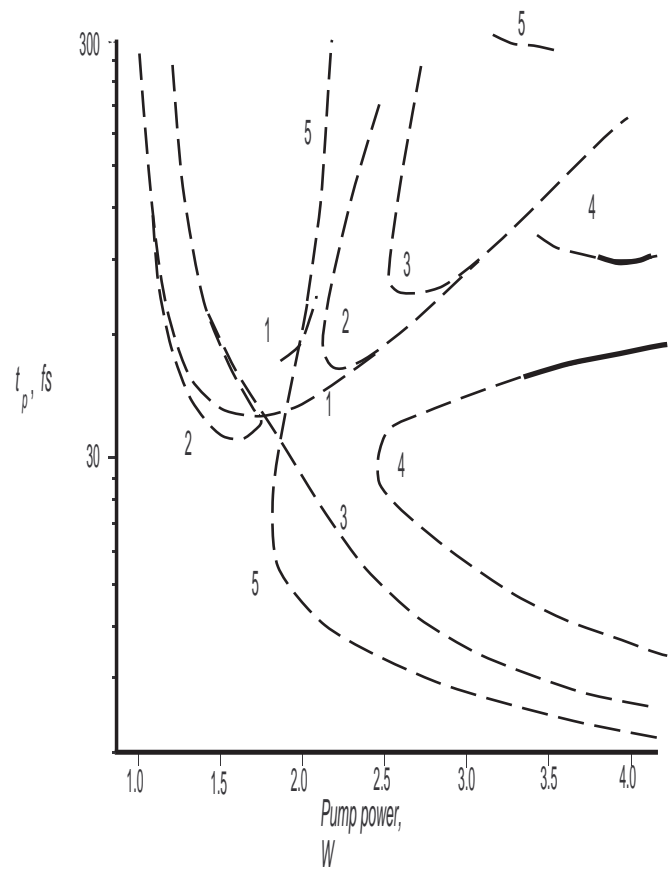}
	\end{center}

	\caption{Pulse duration $t_{p}$ versus pump power}
\end{figure}

\begin{figure}
	\begin{center}
		\includegraphics{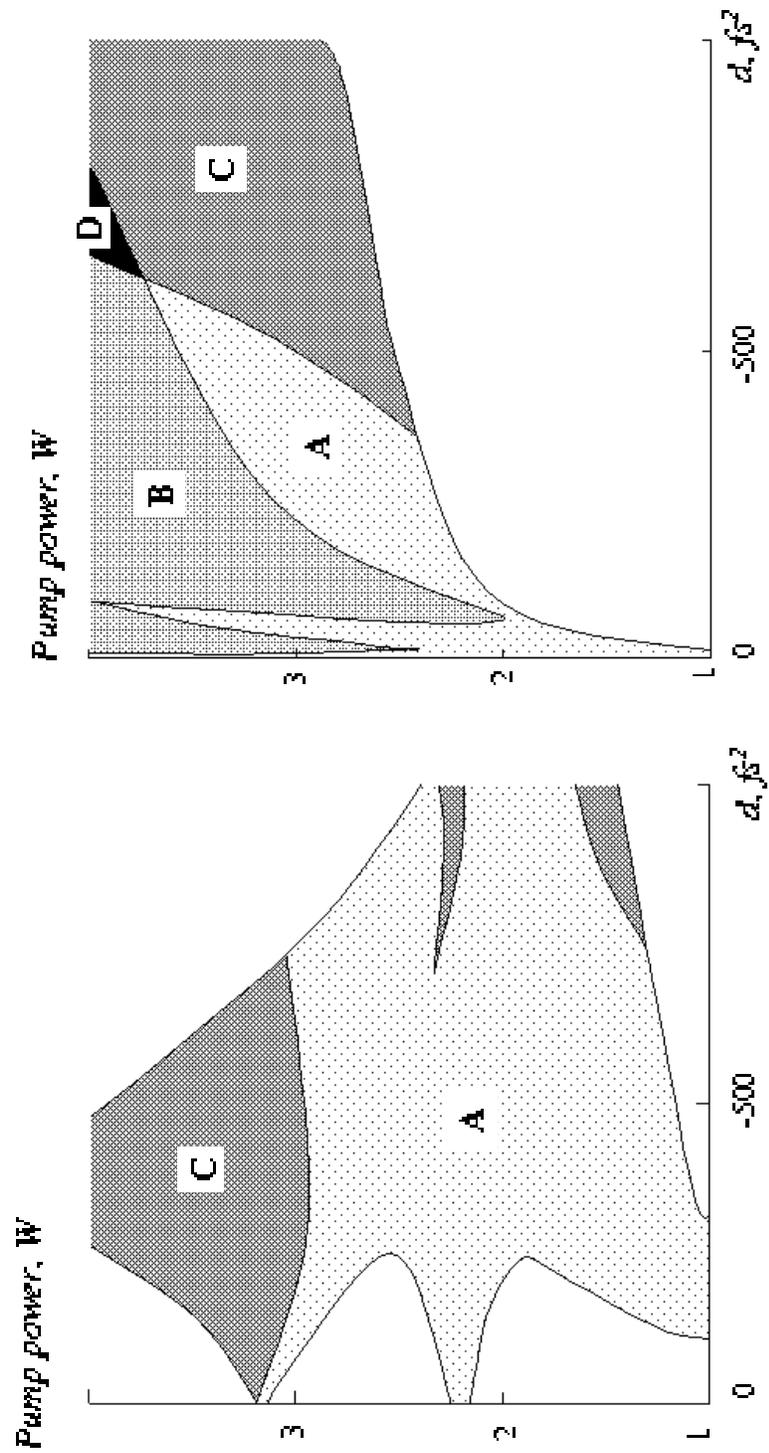}
	\end{center}

	\caption{Regions of pulse existence}
\end{figure}

\begin{figure}
	\begin{center}
		\includegraphics{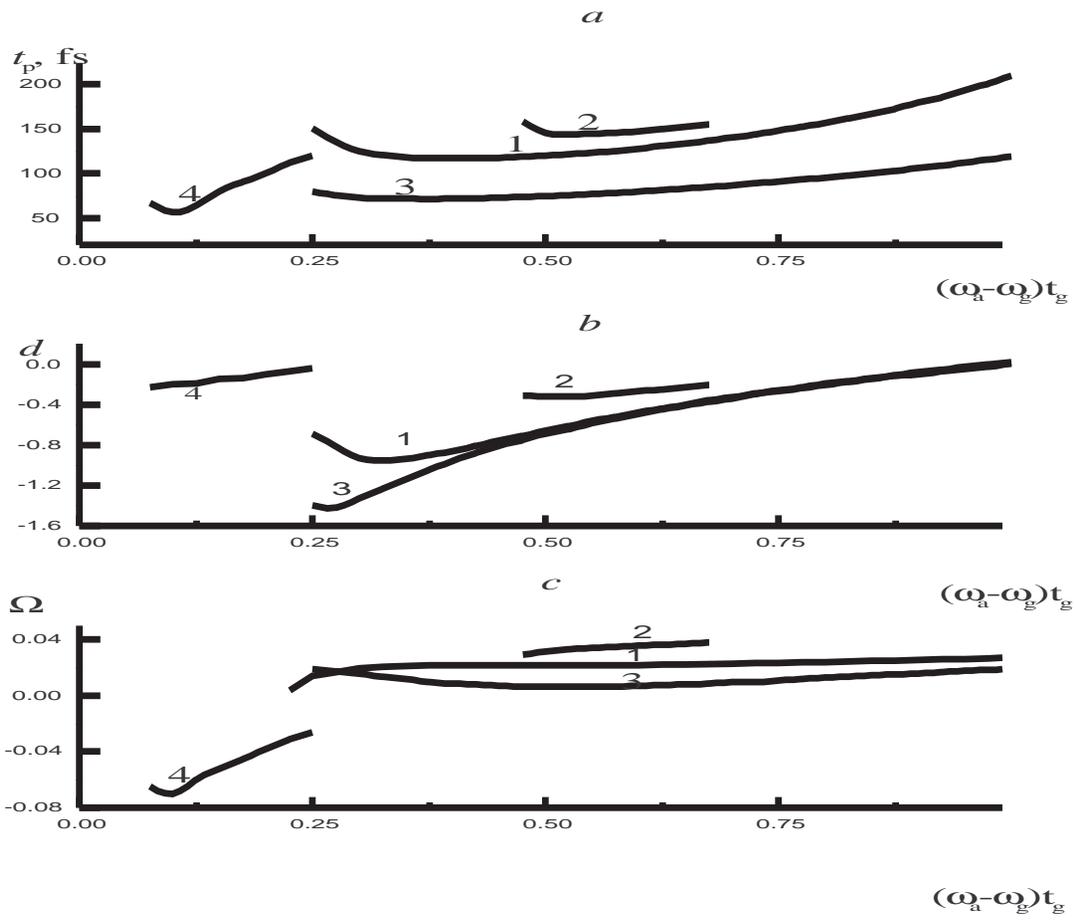}
	\end{center}

	\caption{$t_{p}$ (\textit{a}), $d$ (\textit{b}) and
$\Omega t_{g}$ (%
\textit{c}) for chirp-free solution}
\end{figure}

\begin{figure}
	\begin{center}
		\includegraphics{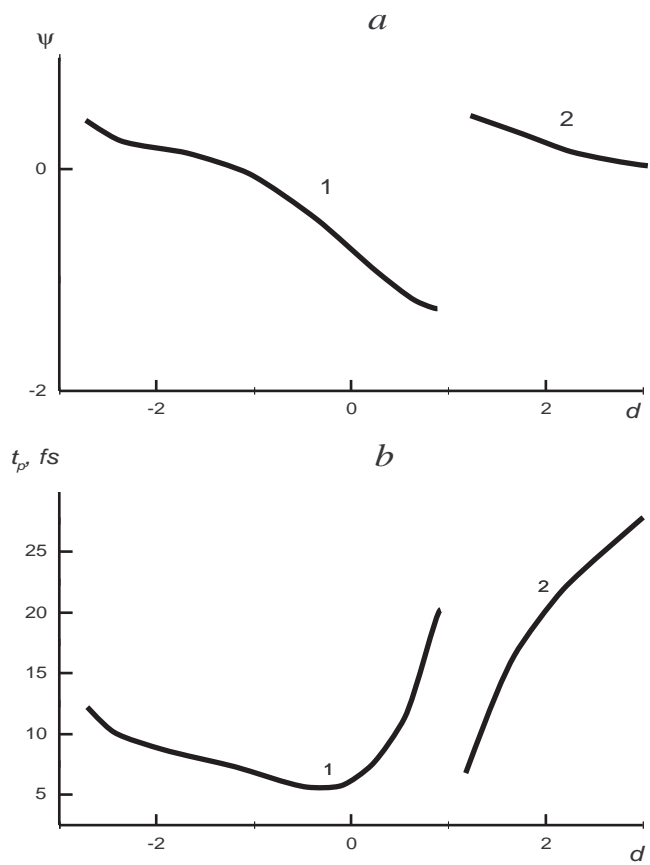}
	\end{center}

	\caption{$\psi $ (\textit{a}) and $t_{p}$ (%
\textit{b}) versus $d$}
\end{figure}

\begin{figure}
	\begin{center}
		\includegraphics{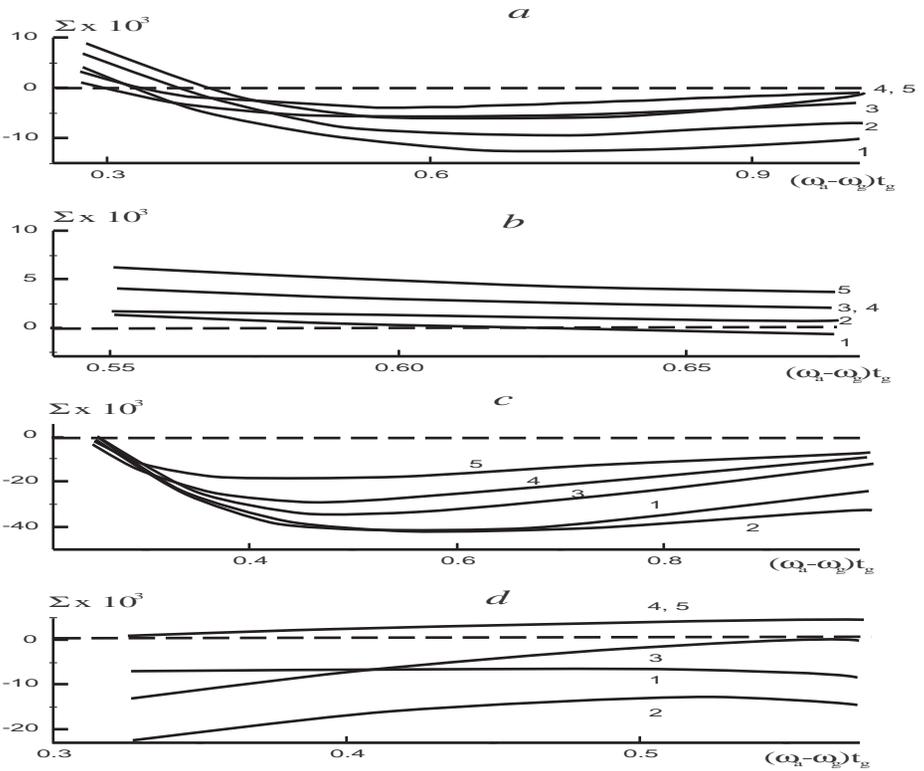}
	\end{center}

	\caption{Net-gain $\Sigma $ behind the pulse tail}
\end{figure}

\begin{figure}
	\begin{center}
		\includegraphics{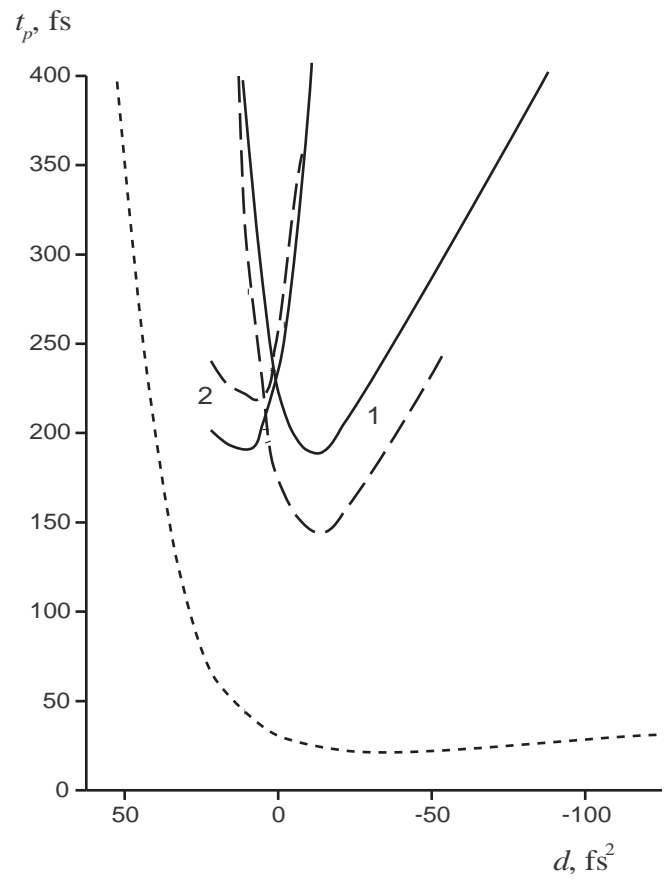}
	\end{center}

	\caption{Duration of the stable pulse $t_{p}$ versus GVD $d$}
\end{figure}

\begin{figure}
	\begin{center}
		\includegraphics{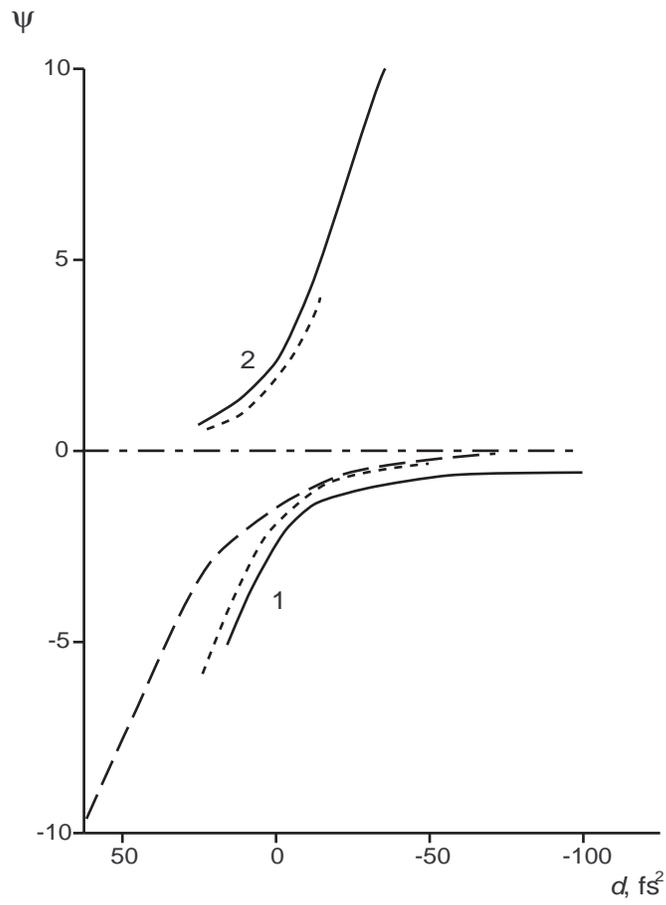}
	\end{center}

	\caption{Chirp $\psi $ of the stable pulse versus GVD $d$}
\end{figure}

\begin{figure}
	\begin{center}
		\includegraphics{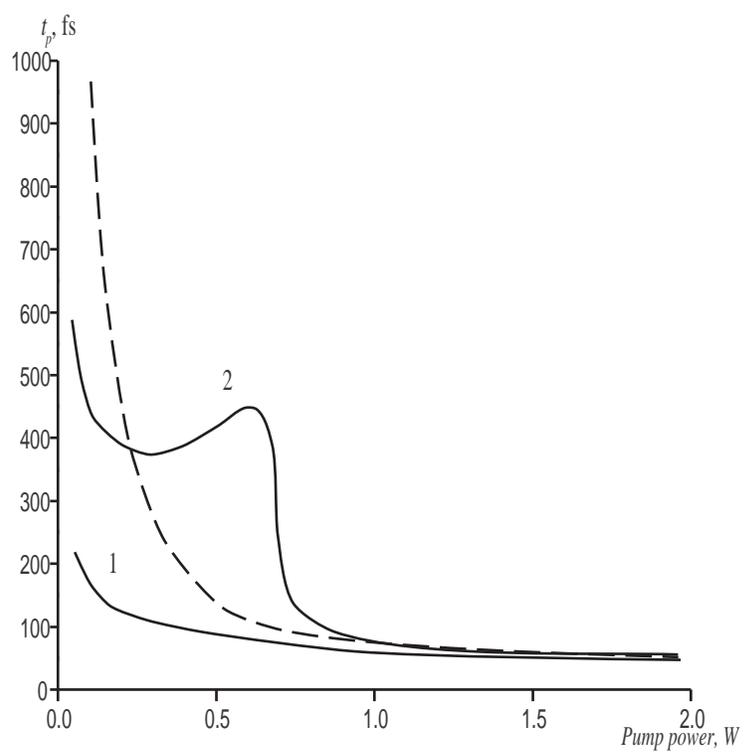}
	\end{center}

	\caption{Duration of the stable pulse $t_{p}$ versus pump power}
\end{figure}

\end{document}